\documentstyle[twocolumn,aps]{revtex}

\begin{document}

\title{Superconducting Gap Nodal Surface and Fermi Surface:\\
their partial overlap in cuprates}

\author{G. Baskaran$^\ast$, D. G. Kanhere$^\dagger$,
        Mihir Arjunwadkar$^\dagger$ and Rahul Basu$^\ast$\cite{email}}

\address{${}^\ast$    The Institute of Mathematical Sciences,
                      Madras 600 113, India}
\address{${}^\dagger$ Department of Physics, University of Poona,
                      Pune 411 007, India}

\date{\today}

\maketitle

\begin{abstract}

Electron correlation in cuprates leads to a global constraint
$\sum_{\bf k} \Delta_{\bf k} = 0$ on the gap function $\Delta_{\bf k}$
resulting in a gap nodal surface.
We give physical arguments supported by numerical results and discuss
some experimental results to argue that correlations also lead to a
local constraint on charge fluctuations in ${\bf k}$-space close to
the Fermi surface, which may result in a substantial overlap of the
Fermi surface with the gap nodal surface.

\end{abstract}


%
%

An actively debated issue in the experimental study of high-$T_c$
cuprate superconductors \cite{Exp,ohsugi,arpes} is the symmetry
and detailed ${\bf k}$-dependence of the gap function $\Delta_{\bf k}$.
We suggest a form of $\Delta_{\bf k}$, based on physical arguments
and numerical results, which is in qualitative agreement with the
available experimental results.
The form that we suggest is a generalised version of an early
suggestion of Anderson \cite{pwa1} in the context of cuprates and
even an earlier suggestion of Cohen \cite{cohen} in the context
of sodium metal.
In our scenerio, what is of primary importance is the vanishing of
gap function on patches of the fermi surface(FS), leading to a
gapless superconducting state with finite density of states for
quasiparticle excitations at the chemical potential.
That is, in the vicinity of finite portions of the FS the gap
function has a non-zero value $\Delta_{\bf k} \approx \Delta_1$
and in the vicinity of the rest of the portions on the FS the
gap function has the form
$\Delta_{\bf k} \approx (\vert {\bf k} \vert - {\bf k}_F) \Delta_2$.
Here $\Delta_1$ and $\Delta_2$ are finite and are weakly dependent
on the wave vector ${\bf k}$.
The issue of the symmetry of the gap function in the context of the
above suggestions is discussed at the end of this paper.

The local constraint on double occupancy in the conducting $CuO_2$
layer of the cuprates has non-trivial consequence on the structure
of the gap function.
In a simple tight binding one band large-$U$ Hubbard model or the
$t-J$ model of the $CuO_2$ layers (for which there is ample
theoretical and experimental evidence) this implies essentially
zero on-site pairing amplitude in the superconducting state:
$\Delta_{ii} \equiv
\langle
        c_{i\uparrow}^\dagger c_{i\downarrow}^\dagger
\rangle = 0$,
which in ${\bf k}$-space means
\begin{equation}
   \Delta_{ii} = \sum \langle
                         c_{{\bf k} \uparrow}^\dagger
                         c_{-{\bf k} \downarrow}^\dagger
                      \rangle = \sum_{\bf k} \Delta_{\bf k} = 0
\end{equation}
assuming usual zero momentum pair condensation.
The above is a global constraint, which forces the gap function
to develop a nodal line (in 2D) or nodal surface (in 3D),
such that the integral of $\Delta_{\bf k}$ on the two sides of
the nodal surface cancel each other.
We call this a gap nodal surface (GNS).

An important question is the shape and location of the gap nodal
surface in ${\bf k}$-space.
A simple possibility is a gap with  $d_{x^2 - y^2}$ symmetry
\cite{dwave} which has nodal lines $k_x = \pm k_y$ and the
global constraint (Eq. (1)) is ensured by the symmetry.
On the other hand, the originial RVB scenerio of Anderson
\cite{pwa1} as well as a simple mean field theory \cite{bza}
for the undoped Mott insulator suggested a gap function
(for the pre-existing neutral Cooper pairs) of the form
$\cos k_x + \cos k_y = 0$, which satisfies the global constraint
in ${\bf k}$-space and also the {\em GNS coincides with the tight
binding FS at half filling}.
Anderson has suggested that on doping the Mott insulator
the GNS will also shrink and continue to overlap with the FS,
thereby making it a truly gapless superconductor.
However, Anderson did not give any physical arguments in support
of his conjecture \cite{pwa1}.

In this letter, we will argue that real space correlation changes
the nature of charge fluctuations in ${\bf k}$-space close to the
FS and it is incompatible with the nature of charge fluctuations
demanded by a finite value of $\Delta_{\bf k}$ on the FS.
This incompatibility reduces the pairing correlations (if possible,
to zero) on the FS.
We also present numerical results on small one and two dimensional
Hubbard and $t-J$ clusters to substantiate the form of our gap function.
Towards the end, we discuss some experimental results including the
recent ARPES \cite{arpes} results in the light of our suggestion.

Let us begin by examining the behaviour of the standard BCS wave
function close to the FS:
\begin{equation}
   \vert \psi_{BCS} \rangle = \prod_{\bf k} (u_{\bf k} + v_{\bf k}
                                  c_{{\bf k} \uparrow}^\dagger
                                  c_{-{\bf k} \downarrow}^\dagger)
                                  \vert 0 \rangle
\end{equation}
which may be rewritten as
\begin{equation}
   \prod_{\bf k}{}^\prime (u_{\bf k}^2 +
      \sqrt{2} u_{\bf k} v_{\bf k} b_{{\bf k},-{\bf k}}^\dagger +
       v_{\bf k}^2 b_{{\bf k},-{\bf k}}^\dagger b_{{\bf k},
       -{\bf k}}^\dagger) \vert 0 \rangle,
   \label{bcs}
\end{equation}
where the product is over only half of the ${\bf k}$-space
(e.g. $k_x > 0$ and all $k_y$ in 2D), with
$ b_{{\bf k},-{\bf k}}^\dagger = \frac{1}{\sqrt{2}} (
             c_{{\bf k}\uparrow}^\dagger c_{-{\bf k}\downarrow}^\dagger
           - c_{{\bf k}\downarrow}^\dagger c_{-{\bf k}\uparrow}^\dagger )
$
as the singlet pair creation operator on points ${\bf k}$ and
$-{\bf k}$ in ${\bf k}$-space.
It is clear that $u_{\bf k}^2$ is the {\em probability amplitude}
of finding zero singlet pair with momentum $({\bf k},-{\bf k})$,
${\sqrt 2} u_{\bf k} v_{\bf k}$ is that of finding one singlet pair
(of charge $2e$), and $v_{\bf k}^2$ that of finding
two singlet pairs (of total charge $4e$).
The BCS state has identical phase relations for various configurations
of pair occupancy in ${\bf k}$-space.
That is, when the product in Eq. (\ref{bcs}) is expanded out,
the resulting sum has identical phase for all terms, each term
corresponding to different configurations of the $({\bf k},-{\bf k})$
occupancy.
Superconductivity can thus be thought of as a coherent charge-$2e$
fluctuating state in ${\bf k}$-space.
Since $u_{\bf k} v_{\bf k}$ is non-zero only in a thin energy shell
around the FS, the coherent $2e$ charge fluctuation is concentrated
around the FS.
(It is interesting to note that this coherence in ${\bf k}$-space
results in phase coherence among the Cooper pairs in real space also).
Away from the shell, we either have a completely filled band (inside
the FS) or a completely empty band (outside the FS) and hence no
charge fluctuations.
Thus, $u_{\bf k} v_{\bf k} (\sim \Delta_{\bf k})$ is a measure of
coherent charge fluctuations or k-space electron pair compressibility.

We now argue that strong correlations in real space lead to a
suppression of such coherent charge fluctuations close to the FS.
Strongly correlated electrons in 1D and 2D, described by a large-$U$
Hubbard model, have certain unique features close to the FS.
It is well known that in the 1D Hubbard model there is singular forward
scattering between two electrons with opposite spins close to the FS.
This leads to a finite phase shift \cite{pwa2} at the FS and the
consequent failure of the Fermi liquid theory, resulting in the
vanishing of the discontinuity in $n_{\bf k}$ at the FS
(Luttinger liquid behaviour).
It also implies an effective hard-core repulsive pseudopotential
between electrons with opposite spins close to the FS.
Thus no two electrons close to the FS, with opposite spins, can
have the same momentum, thereby making ${\bf k}$-points close to
the FS essentially singly occupied.
Single occupancy in the vicinity of the FS has been seen in numerical
works on one and two dimensional $t-J$ models \cite{rrps}.
Ours is a first and natural explanation of this.
Freezing of occupancy to one at every point in ${\bf k}$-space close
to the FS also means reduced (or vanishing) pair fluctuations on the FS.
Coherent pair fluctuations are thus unlikely to develop on or very close
to the FS, but are not forbidden away from the FS.
It is therefore likely that the GNS, implied by the global constraint
(Eq. (1)) will coincide with the FS.

Generalizations of the above argument to two and higher dimensions is
straightforward provided the projective constraint of no double occupancy
in real space leads to a finite phase shift for forward
scattering on the FS and the consequent failure of Fermi liquid theory.
In particular, the 2D case can also be understood in the spirit of
Anderson's tomographic Luttinger liquid picture \cite{tll} where we
have a collection of 1D chains in ${\bf k}$-space.

%
%

By an exact diagonalisation of finite $t-J$ and Hubbard clusters,
we find the gap function $\Delta_{\bf k}$ by diagonalizing the
two particle reduced density matrix
\begin{equation}
   A_{{\bf k} {\bf k}^\prime} = \langle
                                  b^\dagger_{{\bf k}, -{\bf k}}
                                  b_{{\bf k}^\prime, -{\bf k}^\prime}
                                \rangle
\end{equation}
which has the eigenfunction decomposition
\begin{equation}
   A_{{\bf k} {\bf k}^\prime} = \sum_\alpha \lambda_\alpha
                                     \Delta_\alpha({\bf k})
                                     \Delta^\ast_\alpha({\bf k}^\prime),
\end{equation}
where $\lambda_{\alpha}$ and $\Delta_\alpha({\bf k})$ are the $\alpha$-th
eigenvalue and eigenfunction of the $N \times N$ matrix
$A_{{\bf k} {\bf k}^\prime}$.
Here $N$ is the number of points in the Brillouin zone and the index
$\alpha$ orders $\lambda_{\alpha}$'s as $\lambda_1 \geq \lambda_2 \geq
\lambda_3 \geq \ldots$.
Superconducting ODLRO is signalled \cite{yang} by a macroscopic
separation of the largest eigenvalue $\lambda_1$ from the next one
$\lambda_2$, {\em i. e.} $\lambda_1 - \lambda_2 \approx N$.
The required gap function $\Delta_{\bf k}$ is the eigenvector
$\Delta_1({\bf k})$ corresponding to the largest eigenvalue $\lambda_1$.
For example, for the standard BCS ground state, $\lambda_1 = N$,
$\Delta_1({\bf k}) = u_{\bf k} v_{\bf k} $ and $\lambda_\alpha = 0$,
for $\alpha = 2, 3 \ldots N$.

%
%

We have evaluated $\lambda_{\alpha}$ and $\Delta_\alpha({\bf k})$
using exact diagonalization of 8-site Hubbard and 16-site $t-J$
chains and $\sqrt{8} \times \sqrt{8}$ Hubbard and $4 \times 4$
$t-J$ planes.
We chose a physically relevant range of ${U \over t} \approx 5 - 10$
and ${J \over t} \approx 0.1 - 0.5 $.
For the 16-site $t-J$ chain, we find that $\Delta_{\bf k}$ has a node
at $k = k_s$ (Fig. 1), and within finite size limitations $k_s = k_F$,
as seen from Fig. 2.
In contrast, in a negative-$U$ Hubbard model we find no node,
as expected (fig. 1).
In our opinion this is a strong indication for the GNS to coincide
with the FS in 1D.
In Figure 3, we display the eigenvalue spectrum for the two-particle
density matrix in order to display the separation of the largest
eigenvalue.
For 2D clusters of size $\sqrt{8}\times \sqrt{8}$ and $4 \times 4$
with 2 holes we find that the eigenfunction corresponding to the
largest eigenvalue, $\Delta_1({\bf k})$, has s-symmetry with a GNS
close to the FS, as seen from Fig. 4.
For the $4 \times 4$ case with 2 holes, the second eigenvector
$\Delta_2({\bf k})$, has $d_{x^2-y^2}$ symmetry (Fig. 4).
When we look at the eigenvector corresponding to the $d_{x^2 - y^2}$
symmetry, we also find a suppressed pairing just outside the FS.
This could also imply that in the thermodynamic limit, the local
constraint may also develop GNS even in the d-state.
Note that an earlier work of Riera and Young and a recent work of Ohta
{\em et al.} \cite{num} is not inconsistent with the present numerical
results, and in particular, they were not studying the behaviour of GNS,
which is our main focus.

We now turn to some of the relevant experimental work in this context.
{}From NQR NMR \cite{Exp,ohsugi} studies, it is clear that ${1 \over T_1}$
follows the Korringer law ${1 \over T_1} = kT$ for $T < 10^\circ K$, in
the superconducting state of $La_{2-x}Sr_xCuO_4$ for various dopings,
suggesting a finite density of quasiparticles at the chemical potential.
One of the existing explanations for this is the presence of strong
scatterers (close to the unitarity limit), which in a d-wave
superconducting state will lead to gapless excitations \cite{ohsugi}.
The number of strong scatterers $N_s$ needed to explain the required
density of states at the fermi level is unphysically high
(${N_s \over N} \sim 0.2$!).
The ubiquitous linear specific heat in the superconducting state also
seems to survive even in purer samples with substantial slope \cite{mason}.
For $La_2CuO_4$ neutron scattering also gives an $S(q,\omega)$ which
indicates the presence of gapless excitations in the superconducting state.
Once again we need a large density of strong impurity scatterers to explain
the observed density of states.
Indications for gapless excitations in YBCO is also present in the infrared
conductivity studies \cite{tanaka}.
One distinct possibility, consistent with the proposal that we make in
this letter, is that these gapless excitations are intrinsic and are
arising from the regions of FS where the superconducting gap vanishes
on the FS (but are present inside and outside).
Since there are two parameters in the present scenerio, one, the area over
which the gaplessness on the FS is there and the other, the slope of the
gap across the FS on the GNS, we can make a simple fit to the experiment,
particularly NMR relaxation giving a value of about 30\% of gapless region
on the FS and a value of the slope $\Delta_2 \approx {1 \over 3} v_F$
(when the gap is expressed in terms of fermi energy).
As doping increases from zero, the FS shrinks dragging part of the GNS with it.
(Recall that the GNS coincides with the FS in 2D in our numerical result and
mean field theory at half-filling).
What causes the non-overlap of the GNS with the FS on some parts in the BZ?
One possible explanation is the enhancement of the interlayer pair tunneling
matrix element $\frac{t^2_{\perp}(k)}{t}$ in those directions in $\bf k$-space.
For example, as emphasized by Chakravarty {\em et al.} \cite{chaks},
$\frac{t^2_{\perp}(k)}{t}$ is largest in the $(0,\pi)$ and $(\pi,0)$
directions which enhances pairing in those regions of the FS by keeping
the GNS away.
Our suggestion (Fig. 5) incorporates this idea as well as the eight nodal
points of the recent ARPES data \cite{arpes}.
What we suggest is that the eight points are likely to be nodal lines
which could be resolved with better experimental resolution.
(Fedro and Koelling, Chen and Tremblay \cite{fedro} have suggested
the possibility of several point nodes (in 2D) in an extended s-wave state).
More importantly, it is important to track the GNS whose existence
is demanded by a $t-J$ type of modelling.
(Finer structure in the GNS-FS overlap like possible lumps in the
$(\pi,\pi)$ direction could arise due to residual interactions and
this is incorporated in the figure).

Figure 6 is an attempt to incorporate d-symmetry in our scheme,
although at present neither our numerical results nor any experiment
indicate such a form of the gap function.
Thermodynamic measurements as well as $1 \over T_1$ behavior can
hardly distinguish figures 5 and 6.
However, tunneling measurements can do so and that brings us to the
issue of the symmetry of the gap.
As has been pointed out by Stamp and collaborators \cite{stamp} the
gaplessness on the FS makes many of the properties similar to a gapless
situation like a d-wave superconductor.
Since several tunneling measurements seem to indicate a $d_{x^2-y^2}$
symmetry, we would like to stress that the gaplessness that we have
discussed till now can occur in systems with either s or d symmetry.
However additional point-like nodes (in 2D) as demanded by the symmetry
will occur in a d-wave situation.

In conclusion, we have provided some qualitative arguments with numerical
and experimental support to suggest a partial overlap of the GNS with the FS.
It should be pointed out that our numerical support for both 1D and 2D is
rather good (though better for 1D) and it is important to study the shape of
the GNS with higher experimental accuracy.

%
%

\acknowledgements

We would like to thank V. N. Muthukumar for many useful
discussions.
G. B. wishes to thank G. Aepoli, A. Ramiriez and Z. -Q. Zhang for
discussions and experimental results.
Partial financial assitance was provided under Project No.
SP/S2/M-47/89 by the Department of Science and Technology and DST
project SBR 32 of the National Superconductivity Programme.

%
%

%
%

{\bf Figure Captions}

\begin{enumerate}

\item
Gap function $\Delta_1(k)$ for a 16-site $t-J$ chain at half
filling (curve A, $J = 0.24$) and 2 holes (curves B, $J = 0.08 - 0.32$).
$k_F = {\pi \over 2}, {3\pi \over 8}$ for 0 and 2 holes
respectively.
For comparison, we have also plotted the gap function for an 8-site
Hubbard chain with $U=-5$ (curve C, 2 holes, $k_F={\pi \over 4}$).

\item
$k_s$ vs. $k_F$ for 2, 4, 6 and 8 holes in a 8-site Hubbard chain with
$U = 10$, where $k_s$ is the node of $\Delta_{\bf k}$.
The straight line is the $k_s = k_F$ line as a ``guide to the eye''.

\item
Eigenvalue spectrum of the density matrix for the 16-site $t-J$ chain.
(1) 2 holes, $J=0.24$;
(2) {\em triplet antiparallel} eigenspectrum for 2 holes, $J=0.24$.

\item
The topmost two gap functions $\Delta_1({\bf k})$ and
$\Delta_2({\bf k})$ for a $4 \times 4$ plane with two holes,
J=0.24.
The numbers above and below a square are the coefficients of the
gap function respectively for the top-most (odd-paired) and the next
(d-wave) state.

\item
A suggested form of the GNS with s-symmetry.
The solid line is the FS and the dotted line is the GNS.
The + and - symbols indicate relative
signs of the gap function across the GNS.

\item
A possible form of the GNS incorporating d-wave symmetry.
Conventions are as in Figure 5.

\end{enumerate}

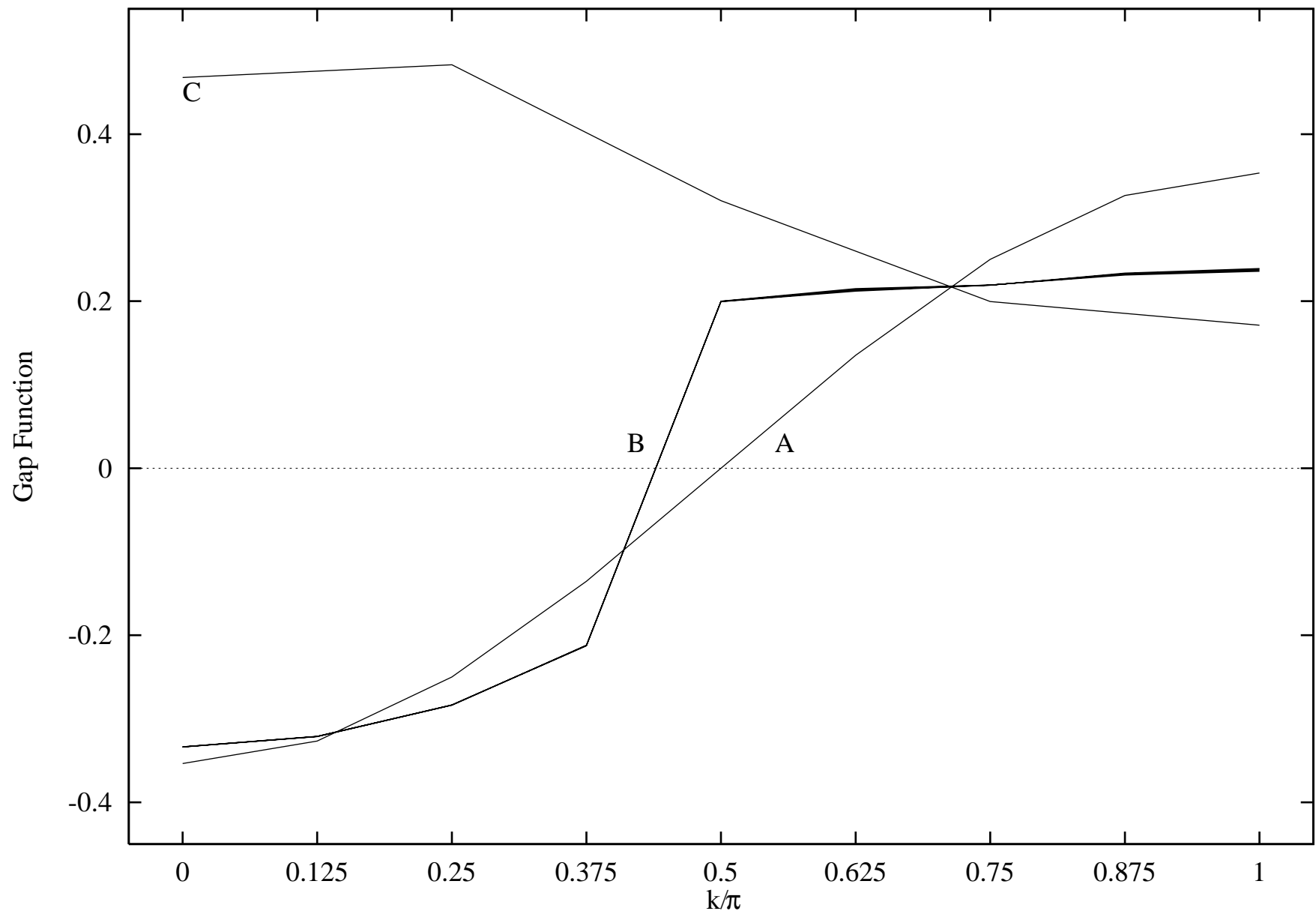
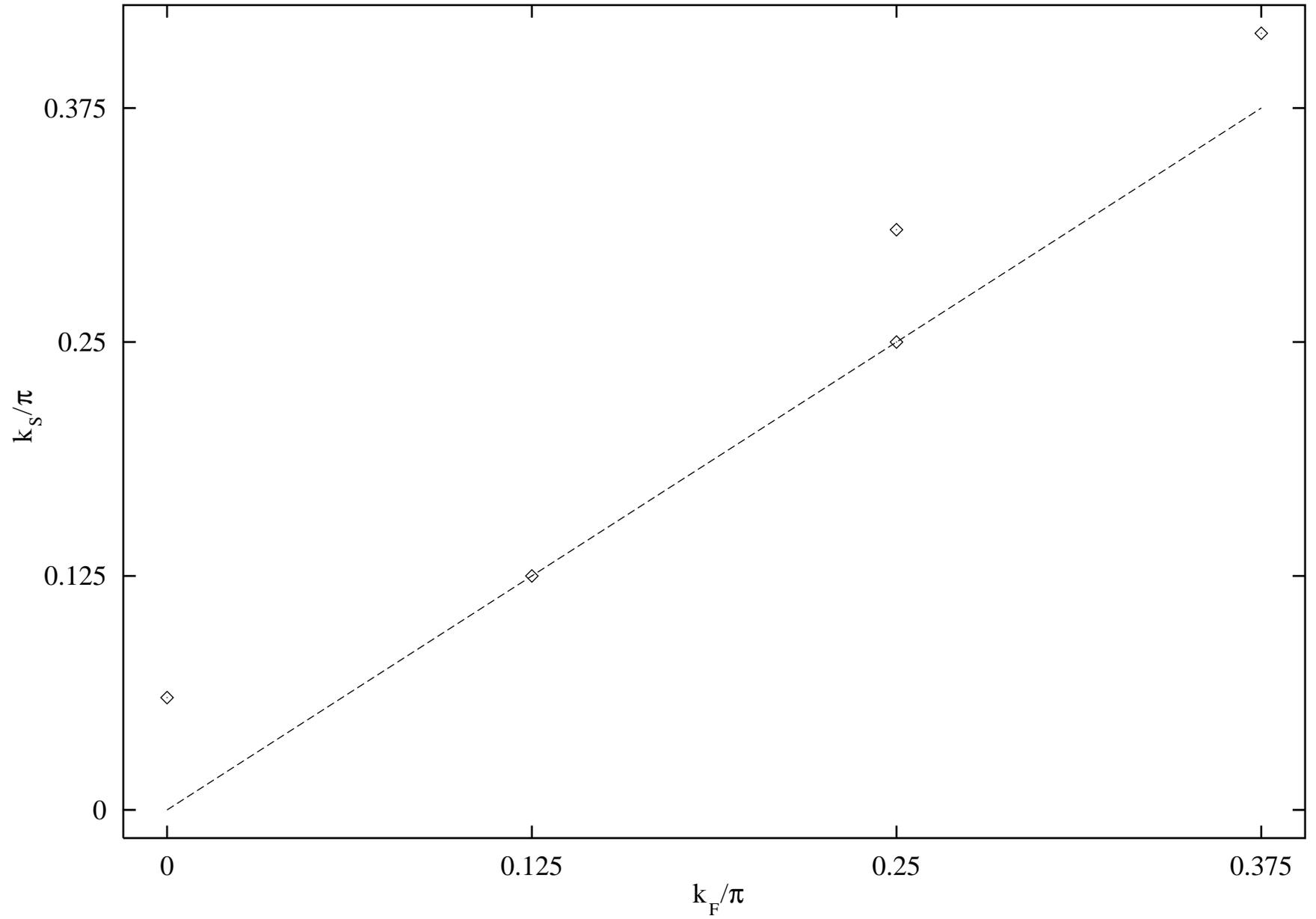
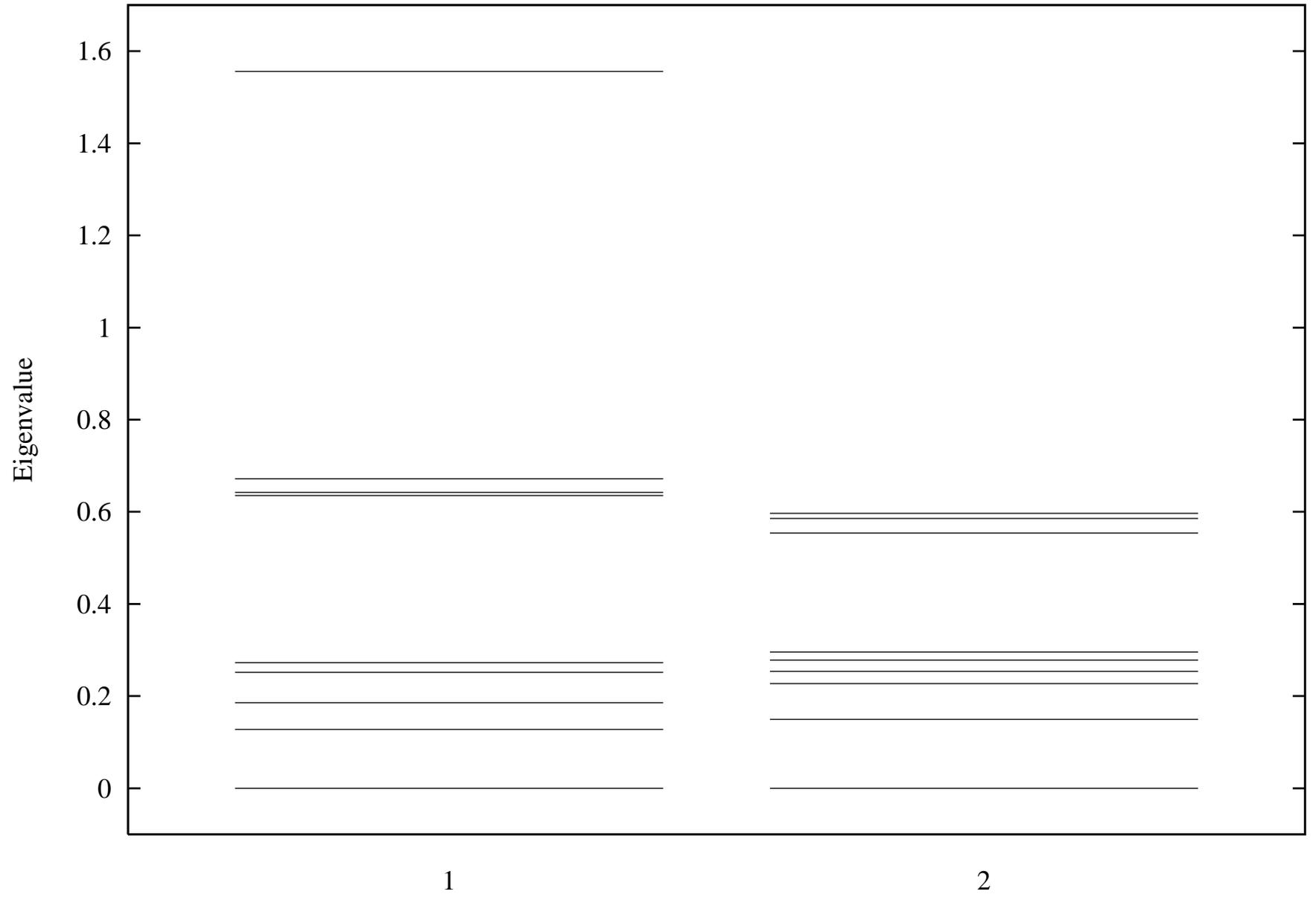
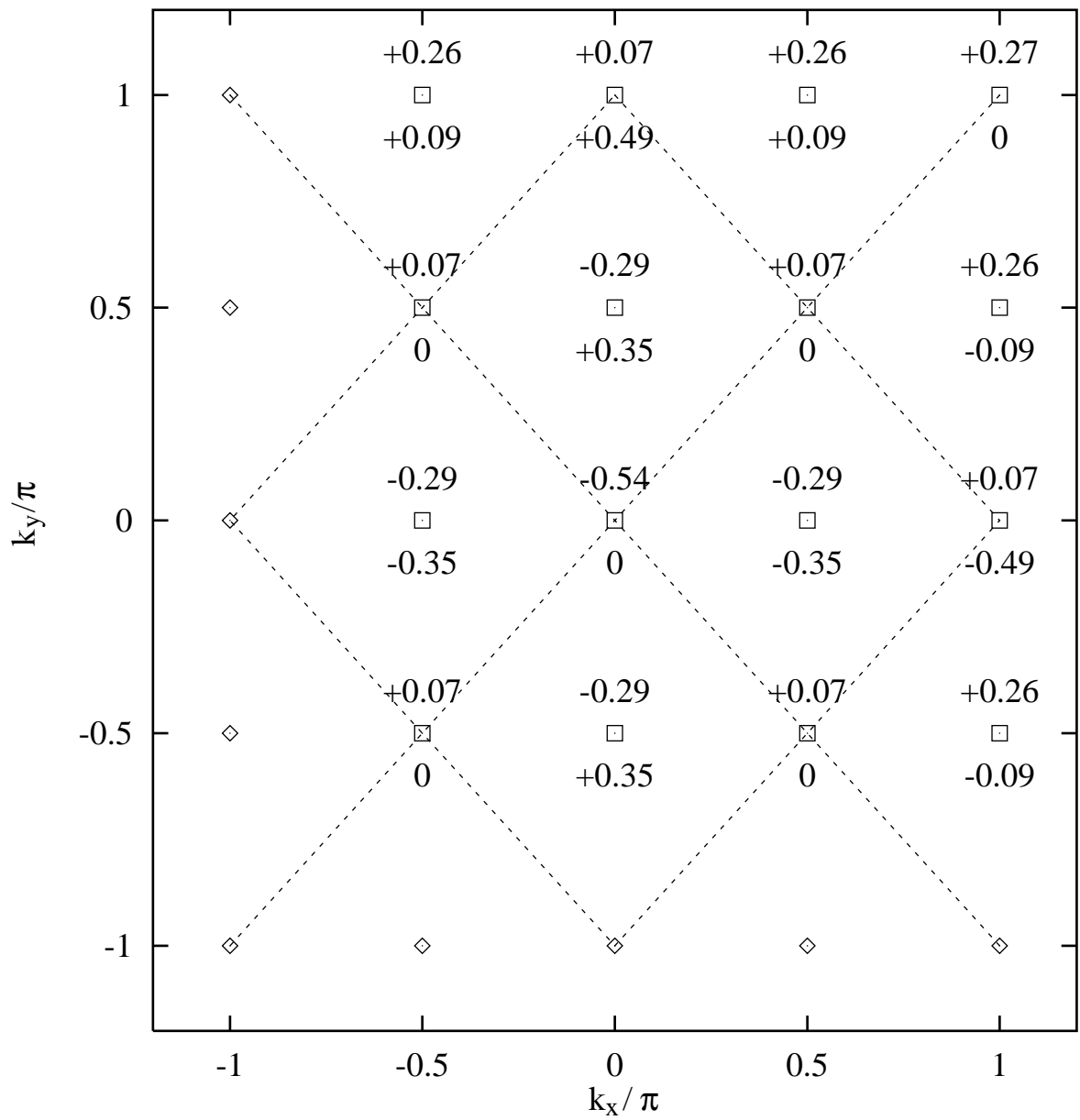
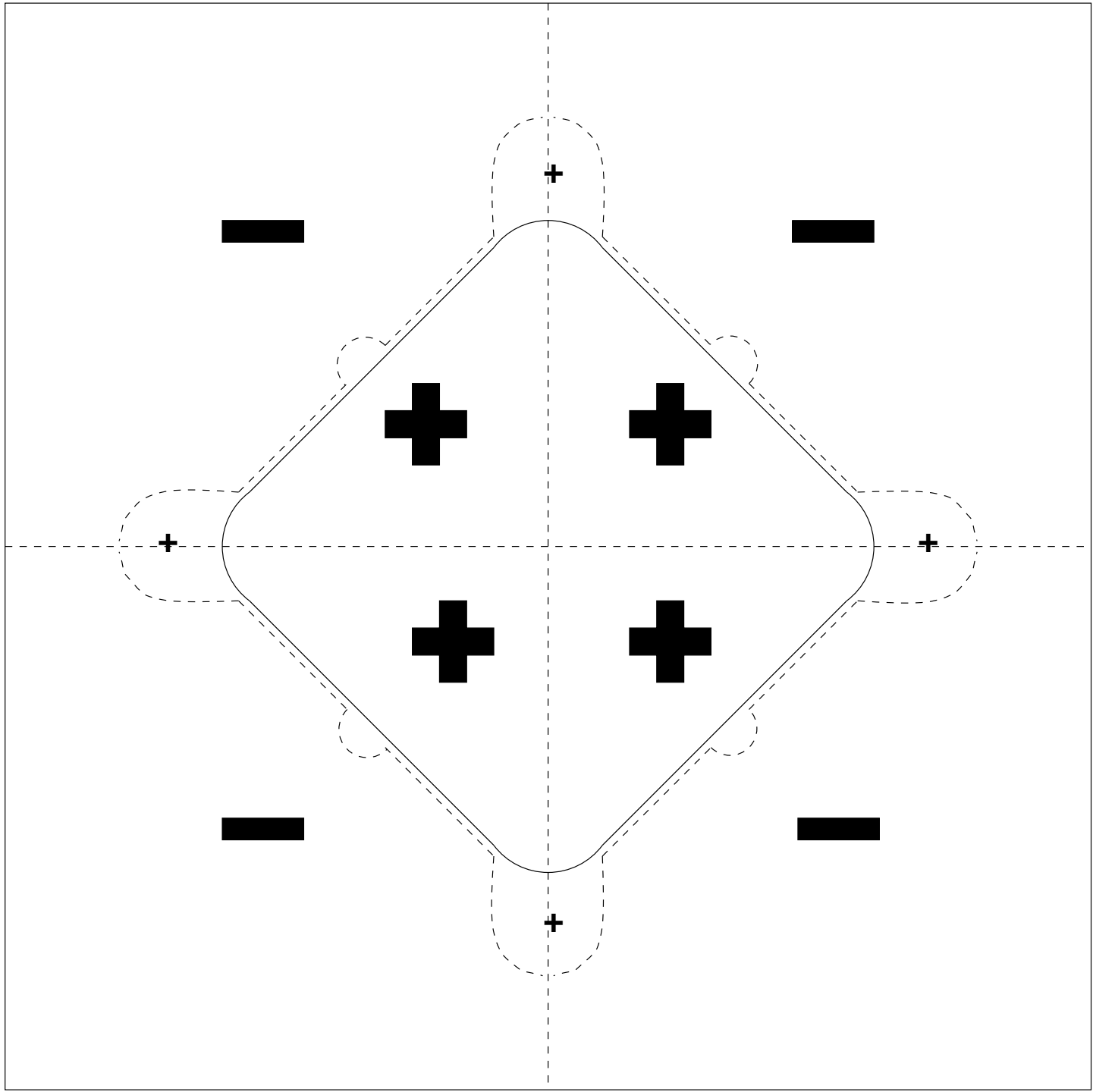
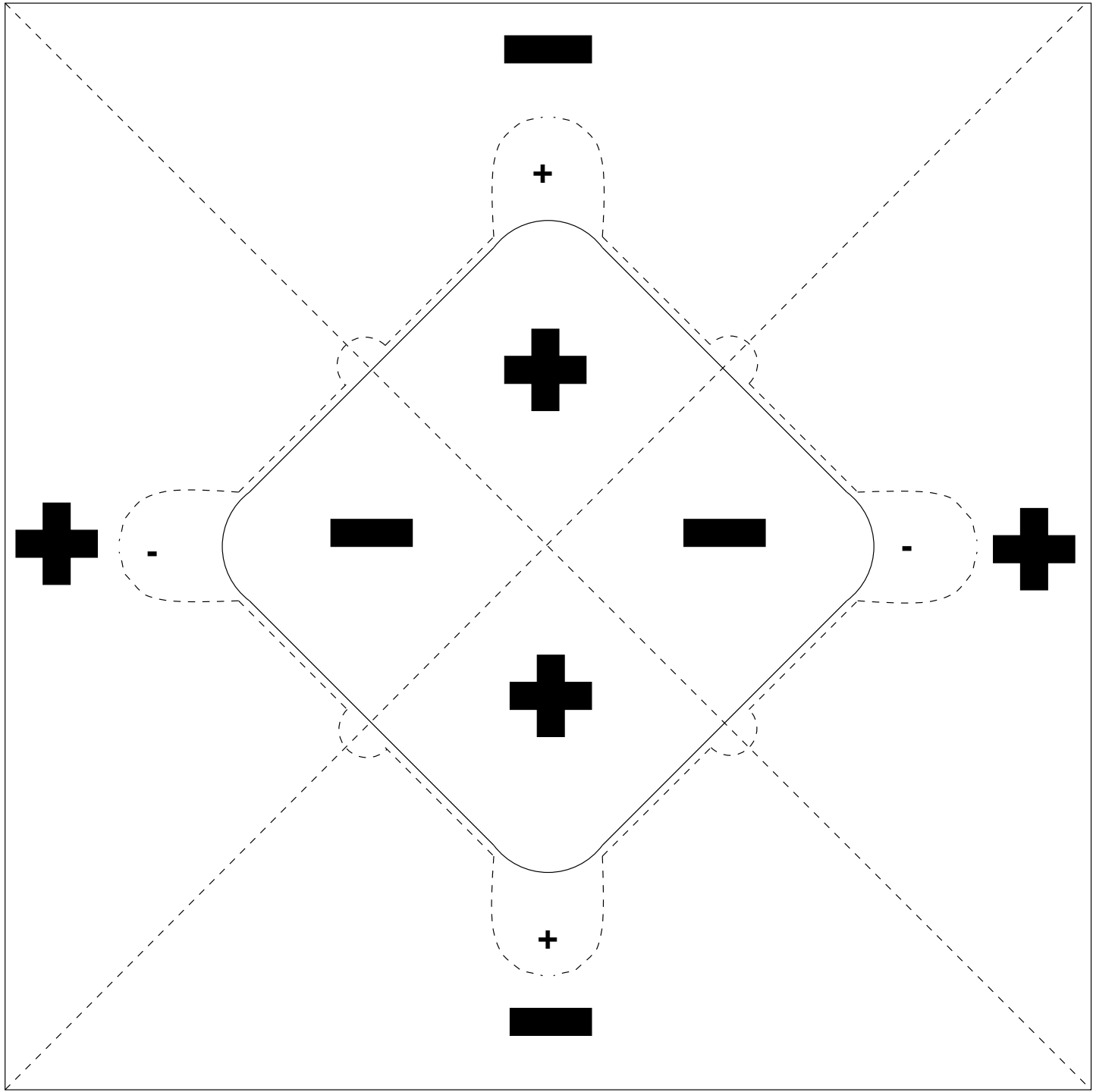
\end{document}